\documentclass[fleqn,usenatbib]{mnras}

\usepackage[T1]{fontenc}

\DeclareRobustCommand{\VAN}[3]{#2}
\let\VANthebibliography\thebibliography
\def\thebibliography{\DeclareRobustCommand{\VAN}[3]{##3}\VANthebibliography}

\usepackage{graphicx}	
\usepackage{amsmath}	
\usepackage{amssymb}	
\usepackage[english]{babel}
\usepackage[normalem]{ulem}

\usepackage[utf8x]{inputenc}
\usepackage{ucs}
\usepackage{microtype}
\usepackage{color}
\usepackage{lettrine}

\usepackage{url}
\usepackage{hyperref}

\usepackage{xspace}
\usepackage{stmaryrd}

\usepackage{empheq}
\usepackage{ifthen}
\usepackage{tensor}
\usepackage{paralist}

\usepackage{amsfonts}
\usepackage{mathrsfs}
\usepackage{extarrows}
\usepackage{verbatim}
\usepackage{upgreek}
\usepackage{wasysym}
\usepackage[bottom]{footmisc}
\usepackage{tgbonum}

\usepackage{natbib}

\usepackage{newtxtext,newtxmath}

\def \be {\begin{equation}}
\def \ee {\end{equation}}
\def \dd {\mathrm{d}}

\def \p {\partial}
\def \l {\left}
\def \r {\right}

\def \bs {\boldsymbol}

\newcommand{\e}[1]{_{\rm #1}}

\newcommand{\beq}{\begin{equation}}
\newcommand{\eeq}{\end{equation}}
\newcommand{\bea}{\begin{eqnarray}}
\newcommand{\eea}{\end{eqnarray}}

\newcommand{\bv}{\textbf{\textrm{v}}}

\newcommand{\bn}{{\bs{n}}}

\newcommand{\HH}{{\cal H}}

\newcommand\ees{\end{eqnarray}}
\newcommand\bees{\begin{eqnarray}}

\definecolor{dgreen}{rgb}{0,0.7,0.0}

\title[Kinematic dipole]{On the kinematic cosmic dipole tension}

\author[C. Dalang and C. Bonvin]{
Charles Dalang\thanks{charles.dalang@unige.ch}
and Camille Bonvin\thanks{camille.bonvin@unige.ch}
\\
Universit\'e de Gen\`eve, D\'epartement de Physique Th\'eorique and Center for Astroparticle Physics, 24 quai Ernest-Ansermet, CH-1211 Gen\`eve 4, Switzerland
}

\date{Accepted XXX. Received YYY; in original form ZZZ}

\pubyear{2021}

\begin{document}
\label{firstpage}
\pagerange{\pageref{firstpage}--\pageref{lastpage}}
\maketitle

\begin{abstract}
Our motion through the Universe generates
a dipole in the temperature anisotropies of the Cosmic Microwave Background (CMB) and also in the angular distribution of sources. 
If the cosmological principle is valid, these two dipoles are directly linked, such that the amplitude of one determines that of the other. However, it is a longstanding problem that number counts of radio sources and of quasars at low and intermediate redshifts exhibit a dipole that is well aligned with that of the CMB but with about twice the expected amplitude, leading to a tension reaching up to $4.9 \sigma$. In this paper, we revisit the theoretical derivation of the dipole in the sources number counts, explicitly accounting for the redshift evolution of the population of sources. We argue that if the spectral index and magnification bias of the sources vary with redshift, the standard theoretical description of the dipole may be inaccurate. We provide an alternative expression which does not depend on the spectral index, but instead on the time evolution of the population of sources. We then determine the values that this evolution rate should have in order to remove the tension with the CMB dipole.
\end{abstract}




\lettrine{M}{easurements} of the Cosmic Microwave Background Radiation by Planck \citep{Aghanim:2018eyx} have confirmed the validity of the $\Lambda$CDM model to an impressive precision. In parallel, measurements of galaxy distributions and sizes from large-scale structure surveys, like BOSS~\citep{BOSS:2016wmc}, the WiggleZ Dark Energy Survey~\citep{2011}, CFHTlenS~\citep{2014}, KiDS~\citep{2021} and the Dark Energy Survey~\citep{DES:2021wwk} are all in agreement with this same model.  
However, despite its success to fit distinct data sets at widely different redshifts, the $\Lambda$CDM model has been seriously put to questions over the years due to the various tensions that have appeared \citep{Riess:2020fzl,Wong:2019kwg,Secrest:2020has}. Although these tensions represent potentially exciting news for the whole physics community, precise calculations of the expected discrepancies must be performed before any claim can be made on the discovery of new physics. This turns out to be a particularly demanding task for most observables: on one hand because General Relativity, which is a non-linear theory, is opaque to perturbation theory beyond certain regimes, and on the other hand because observations are affected by relativistic effects, that distort our image of the sky~\citep{Yoo:2009au,Bonvin_2011,Challinor_2011}. 

In this paper, we focus on one such discrepancy: the mismatch between the dipole observed in the Cosmic Microwave Background (CMB) and the one measured in the distribution of radio sources and quasars. Measurements of the temperature anisotropies of the CMB show a dipole with an amplitude which is two orders of magnitude larger than for other multipoles. If assumed to be entirely due to a Doppler shift coming from our motion (i.e.\ the motion of the Solar System) with respect to the surface of last scattering of the CMB photons, this dipole indicates that we are moving with a peculiar velocity of amplitude $|\bv_o|= 369.82 \pm 0.11$ km s$^{-1} $ with respect to the CMB rest frame \citep{Aghanim:2018eyx}. Galactic peculiar velocities of this order of magnitude are expected \citep{Bertschinger1991} and observed in the nearby Universe \citep{Strauss:1995fz}. 

At large scales, the rest frame of galaxies is expected to coincide with the CMB rest frame. Our motion with respect to this frame is therefore expected to generate also a kinematic dipole in the observed number counts of sources. This dipole has been predicted by \cite{Ellis1984} to have an amplitude
\begin{align}
\mathcal{D}\e{EB}=[2 + x(1+\alpha)] |\bv_o|\, ,
\label{eq:dipEB}
\end{align}
where $x$ is the slope of the integral source counts at the flux limit (often called magnification bias parameter) and $\alpha$ is the spectral index of the population of sources. Note that we work in units such that $c=1$. Measurements of the number counts dipole have been performed in various surveys using millions of sources~\citep[e.g.][]{Colin:2017juj,Bengaly_2018,Secrest:2020has,Siewert:2020krp}. While the direction of this dipole is well aligned with that of the CMB, the observed amplitude is several times larger than what is expected from Eq.~\eqref{eq:dipEB}, if $|\bv_o|$ is the one measured from the CMB dipole. Due to the tight error bars in both the CMB and the number counts measurements, this leads to a non-negligible tension between these two results. For example, \citet{Secrest:2020has} reports a $4.9\sigma$ tension.
 
In this paper, we revisit the theoretical derivation of the kinematic dipole in the number counts of sources. Observationally, the dipole is measured by summing over all sources in the catalog, independently of their distance from the observer. The theoretical dipole varies however with distance and needs to be integrated over the sources distribution to reproduce the observed dipole. This dependence on distance is somewhat hidden in Eq.~\eqref{eq:dipEB} by the fact that $x$ and $\alpha$ are effective parameters averaged over the sources distribution. Here we re-derive the dipole, explicitly accounting for the dependence on distance. We show that the integral over the sources distribution can be done in two different ways: either by integrating over the comoving distance of sources, $r=\eta_0-\eta$, where $\eta$ denotes conformal time; or by integrating over the sources redshift $z$. We show that the two methods are mathematically equivalent, as they should be and that they reduce to Eq.~\eqref{eq:dipEB} in the case where $\alpha$ and $x$ are redshift-independent. 

The advantage of having two different theoretical expressions is that these expressions depend differently on the intrinsic properties of the population of sources: the integration over comoving distance requires knowledge of the parameter $\alpha$, which describes the spectrum of the sources. The integration over redshift on the other hand requires knowledge of the time evolution of the population of sources. Since these parameters are measured differently and rely on different assumptions about the properties of the sources, it is interesting to use both theoretical expressions independently. In particular, we can ask ourselves the question: how should the population of quasars used e.g.\,in~\citet{Secrest:2020has} evolve to reproduce the observed dipole and resolve the tension with the CMB dipole? We compare this with measurements of the evolution of quasars in the spectroscopic survey eBOSS~\citep{Wang:2020ibf} and we argue that a reasonable evolution of sources can remove the tension with the CMB dipole. This of course does not mean that the observed tension is wrong, since different samples can have different evolutions. Nevertheless, our work provides an alternative theoretical expression which can be used to study the robustness of the tension.

The rest of the paper is structured as follow: in Section~\ref{sec:theory}, we calculate the kinematic dipole as a function of comoving distance and as a function of redshift. In Section~\ref{sec:comp}, we show that these two expressions are mathematically equivalent once integrated over the distribution of sources. In Section~\ref{sec:example}, we apply our expression to a concrete example and we determine how the population of sources should evolve with redshift to solve the tension with the CMB. We conclude in Section~\ref{sec:conclusion}.

\section{The theoretical dipole}
\label{sec:theory}

The dipole in the number counts of sources (galaxies or quasars) is a function of the distance between the observer and the sources. If no measurement of the distance is available, one measures the dipole averaged over all sources at different distances. From a theoretical point of view, there are different ways of calculating such a dipole: one can integrate the dipole over the comoving distance of the sources $r\equiv\eta_o-\eta$, over their redshift $z$, or over any other indicator of distance. Since different distance indicators are affected in different ways by the observer's velocity, we expect the resulting distance-dependent dipole to differ. However, once averaged over all sources, all results should be equivalent.

Here we present two derivations: in Section~\ref{sec:distance} we calculate the dipole integrated over the comoving distance of the sources, and in Section~\ref{sec:redshift} we calculate the dipole integrated over the redshift of the sources. Even though the first result is well-known, we present the calculation in detail: this allows us to emphasise the difference between the two methods, but also to make the link with previous relativistic calculations of the redshift-dependent dipole~\citep{Challinor_2011, Maartens:2017qoa,Nadolny:2021hti}.

\subsection{Dipole integrated over comoving distance}
\label{sec:distance}
 
We denote by
\begin{align}
\frac{\dd N}{\dd \Omega \dd r}\big(r, \bn, S>S_*(\nu_o)\big) \label{eq:dNdr}
\end{align}
the number of sources per solid angle $\dd \Omega$ and bin $\dd r$, detected in direction $\bs{n}$, at comoving distance $r$, and with a flux density $S$ larger than $S_*$ in some frequency band $[\nu_o, \nu_o+\dd\nu_o]$. The dipole averaged over all distances $r$ is obtained from the difference between Eq.~\eqref{eq:dNdr} and its angular average, integrated over $r$
\begin{align}
&\Delta^r(\bn)\equiv\mathcal{D}_{\rm kin}^r\cdot \bn\cdot\hat{\bv}_o\label{eq:fracdr}\\
&=\frac{\int_0^\infty \dd r \left[\frac{\dd N}{\dd \Omega \dd r}\big(r, \bn, S>S_*(\nu_o)\big)-\frac{\dd \bar N}{\dd \Omega \dd r}\big(r, S>S_*(\nu_o)\big)\right]}{\int_0^\infty \dd r\frac{\dd \bar N}{\dd \Omega \dd r}\big(r, S>S_*(\nu_o)\big)}\, , \nonumber
\end{align}
where $\hat{\bv}_o=\bv_o/|\bv_o|$ and the angular average over the $2$-sphere $S_2$ is given by
\begin{align}
\frac{\dd \bar N}{\dd \Omega \dd r}\big(r, S>S_*(\nu_o)\big) \equiv \frac{1}{4\pi}\int_{S_2} \dd \Omega \frac{\dd N}{\dd \Omega \dd r}\big(r, \bn, S>S_*(\nu_o)\big)\, . \label{eq:av}
\end{align}
The difference between the direction-dependent number of sources and the angular average which appears in Eq.~\eqref{eq:fracdr} is not only affected by the observer's velocity, but also by intrinsic inhomogeneities in the distribution of the sources, by the peculiar velocity of the sources, and by the so-called relativistic distortions that modify the observed clustering of galaxies. These contributions have been calculated at linear order in perturbation theory~\citep{Yoo:2009au,Bonvin_2011,Challinor_2011}, and can be numerically computed using for example the public codes CLASS~\citep{Blas:2011rf} or CAMB~\citep{Challinor:2011bk}. Since at linear order these termes can simply be added to the kinematic dipole generated by the observer's velocity, we do not include them in our derivation, but simply add them at the end when we apply our formalism to two concrete examples. In what follows we denote by a bar the quantities associated with an observer at rest with respect to the CMB frame, and by a $\delta$ the perturbations generated by the observer's velocity.

A source sitting at position $(r,\bn)$ with observed flux density $S>S_*(\nu_o)$, has an intrinsic luminosity density $L>L_*(r,\bn,\nu_s)$. We can therefore rewrite Eq.~\eqref{eq:dNdr} as
\begin{align}
\frac{\dd N}{\dd \Omega \dd r}\big(r, \bn, S>S_*(\nu_o)\big)&=\frac{\dd N}{\dd \Omega \dd r}\big(r, \bn, L>L_*(r,\bn,\nu_s)\big)\, .\label{eq:dN_L}
\end{align}
The right-hand side of Eq.~\eqref{eq:dN_L} differs from the angular average~\eqref{eq:av} for two reasons: first the fact that $\dd \Omega$ is affected by aberration
\begin{align}
\dd \Omega \equiv \dd \bar\Omega\left(1+\frac{\delta \Omega}{\dd \bar\Omega}\right)=\dd \bar\Omega(1-2 \bn\cdot\bv_o)\, ;    
\end{align}
and second, the fact that the luminosity threshold $L_*(r,\bn,\nu_s)$ depends on direction for a moving observer. 

To calculate the dipole, we split the luminosity threshold into a direction-independent part (corresponding to the luminosity threshold that an observer at rest would infer), and a perturbation
\begin{align}
L_*(r,\bn,\nu_s)=\bar L_*(r, \bar \nu_s)+\delta L_*(r,\bn,\nu_s)\, .
\end{align}
Here $\nu_s=(1+z)\nu_o$ and $\bar \nu_s=(1+\bar z)\nu_o$, with $\bar z$ the redshift for an observer at rest. Since we are interested in the dipolar contribution at linear order in the observer's velocity, we can Taylor expand Eq.~\eqref{eq:dN_L} around $\bar L_*$
\begin{align}
&\frac{\dd N}{\dd \Omega \dd r}(r, \bn, S>S_*)\simeq\frac{\dd N}{\dd\bar \Omega \dd r}\big(r,  L>\bar L_*(r,\bar\nu_s)\big)\left(1-\frac{\delta\Omega}{\dd \bar\Omega}\right)\nonumber\\
&+\frac{\partial}{\partial {L_*}}\left(\frac{\dd \bar N}{\dd \Omega \dd r} (r,L>L_*)\right)\Big|_{L_*=\bar L_*} \cdot \delta L_*(r,\bn,\nu_s)\, . \label{eq:dNdrTaylor}
\end{align}
We see that two effects modify the observed number of sources for an observer in motion: aberration (in the first line); and luminosity fluctuations (in the second line), which account for the fact that a fixed observed flux threshold corresponds to different luminosity thresholds in different directions. 

The luminosity density is related to the flux density by
\begin{align}
L_*(r,\bn,\nu_s)=4\pi \frac{d_L^2(r,\bn)}{1+z(r,\bn)}S_*(\nu_o)\, , \label{eq:luminosity}
\end{align}
where $d_L$ is the luminosity distance (see Appendix~\ref{app:flux} for a derivation of this relation). For an observer in motion, both the redshift $z$ and the luminosity distance $d_L$ depend on direction. Splitting $d_L$ and $z$ into a background part and a perturbation due to the observer's velocity, and using that $\nu_s=(1+z)\nu_o$ also depends on the observer's velocity, we find at linear order
\begin{align}
\delta L_*(r,\bn,\nu_s)=\bar L_*\left(2\frac{\delta d_L}{\bar d_L}-\frac{\delta z}{1+\bar z} \right)-\frac{\partial}{\partial \nu_s}\bar L_*(r,\nu_s)\Big|_{\nu_s=\bar\nu_s} \!\!\cdot\delta\nu_s\, .   \label{eq:dLr}
\end{align}
The last term accounts for the fact that an observer in motion, who observes photons at fixed frequency $\nu_o$, sees different parts of the source spectrum  at fixed distance $r$, in different directions. If the intrinsic luminosity is independent of frequency, this last term vanishes. However, if the spectrum is not flat, this last term modulates the observed number of sources above a given flux threshold.

The link with the standard calculation of Ellis and Baldwin in Eq.~\eqref{eq:dipEB} can be made by using that for radio sources and for quasars the luminosity spectrum obeys $L\propto \nu_s^{-\alpha}$ such that
\begin{align}
\frac{\partial}{\partial \nu_s}\bar L_*(r,\nu_s)\Big|_{\nu_s=\bar\nu_s}=-\alpha\frac{\bar L_*}{\bar \nu_s} \, , 
\label{eq:Lalpha}
\end{align}
and the average number counts scales as
\begin{align}
\frac{\dd \bar N}{\dd \Omega \dd r} (r, S>S_*)\propto S_*^{-x(r)}\, . \label{eq:defx}  
\end{align}
Note that the parameter $x$ is related to the so-called magnification bias parameter, $s$, defined as
\begin{align}
s(z)& \equiv \frac{\p}{\p M_*} \log_{10} \l(\frac{\dd \bar N}{\dd V}(z,M<M_*)\r)\label{eq:s(z)}\\
&=-\frac{2}{5}\frac{\partial \ln\left(\frac{\dd \bar N}{\dd \Omega \dd z}(z, S>S_*) \right)}{\partial \ln S_*}\nonumber\\
& = -\frac{2}{5}\frac{\partial \ln\left(\frac{\dd \bar N}{\dd \Omega \dd r}(r(z), S>S_*) \right)}{\partial \ln S_*}\nonumber\\
&=\frac{2}{5}x(r(z))\, ,\nonumber
\end{align}
where $M$ denotes the magnitude of the sources and $\dd V$ denotes a comoving volume (see Eq.\,\eqref{eq:Comoving_Volume}).
In the third equality, we have neglected the difference between the number of galaxies at fixed redshift and the number of galaxies at fixed distance, since this difference is linear in the observer velocity, and it would lead to a correction quadratic in the velocity in Eq.~\eqref{eq:dNdrTaylor}. We therefore have $x(r(z))=x(z)$.
With this we obtain
\begin{align}
\frac{\partial}{\partial {L_*}}\left(\frac{\dd \bar N}{\dd \Omega \dd r} (r,L>L_*)\right)\Bigg|_{L_*=\bar L_*}=-\frac{x}{\bar L_*} \frac{\dd \bar N}{\dd \Omega \dd r} (S>S_*)\, .
\label{eq:xterm}
\end{align}
Using that 
\begin{align}
\frac{\delta z}{1+\bar z}=\frac{\delta\nu_s}{\bar \nu_s}=-\bn\cdot\bv_o\, ,   \end{align}
and that the perturbation of the luminosity distance at fixed distance $r$ is given by~\citep{Bonvin:2005ps} 
\begin{align}
\frac{\delta d_L(r,\bn)}{\bar d_L(r)}=-\bn\cdot\bv_o\, ,
\label{eq:deltadLr}
\end{align}
we obtain
\begin{align}
\frac{\dd N}{\dd \Omega \dd r}(r, \bn, &S>S_*)=\frac{\dd N}{\dd \bar \Omega \dd r}\big(r,  L>\bar L_*(r,\bar\nu_s)\big)\label{eq:dNchi}\\
&\times\Big\{1+\big[2+x(r)(1+\alpha(r))\big]\bn\cdot\bv_o \Big\}\nonumber\, .
\end{align}
Subtracting the angular average \eqref{eq:av} and integrating over distance $r$ gives rise to the dipole amplitude
\begin{align}
\mathcal{D}_{\rm kin}^r=\int_0^\infty \dd r f(r) \big[2+x(r)(1+\alpha(r))\big] |\bv_o|\, ,  \label{eq:Dchi} 
\end{align}
where 
\begin{align}
f(r)\equiv\frac{\frac{\dd N}{\dd \Omega \dd r}(r, S>S_*)}{\int_0^\infty \dd r\frac{\dd N}{\dd \Omega \dd r}(r,S>S_*)}\, ,
\label{eq:fr}
\end{align}
is the distribution of sources.

Eq.~\eqref{eq:Dchi} reduces to Eq.~\eqref{eq:dipEB} if $\alpha$ and $x$ are independent of $r$. This is however not usually the case. If only $\alpha$ depends on $r$ then on recovers Eq.~\eqref{eq:dipEB} with $\alpha$ replaced by
\begin{align}
\alpha_{\rm eff}&\equiv\int_0^\infty \dd r f(r)\alpha(r)\, .
\end{align}
Likewise, if only $x$ depends on $r$ then one recovers Eq.~\eqref{eq:dipEB} with $x$ replaced by
\begin{align}
x_{\rm eff}&\equiv\int_0^\infty \dd r f(r)x(r)\, .
\label{eq:xeff}
\end{align}
If both $\alpha$ and $x$ depend on $r$ however, i.e.\ if the spectrum of the sources and the magnification bias evolve with time, then Eq.~\eqref{eq:Dchi} does not simply reduce to Eq.~\eqref{eq:dipEB}, since 
\be
x_{\rm eff}\cdot \alpha_{\rm eff}\neq \int_0^\infty \dd r f(r) x(r)\alpha(r)\, .
\ee

Typically, a constant value for both $x$ and $\alpha$ is assumed. For example, in~\citet{Siewert:2020krp} the spectral index is fixed to $\alpha=0.75$ and the slope $x$ is fitted from the whole population of sources, independently on their distance. The same procedure is used in~\citet{Tiwari:2013vff}. There, an improved fit is proposed to characterize the integral source count at the flux limit, but this fit also neglects redshift evolution. In~\citet{Secrest:2020has}, the theoretical prediction is drawn from mock samples. More precisely, they measure the flux density of sources, $S(\nu_o)$, and used it to infer $\alpha$. They find that $\alpha$ varies roughly between $0.75$ and $7$, with a distribution following a decaying power law and averaging to $\langle \alpha \rangle = 1.26$ (see their Fig.\ 2). Mock catalogs are then created by assigning to each source a value for $\alpha$ and for $S(\nu_o)$ drawn from the observed distribution.  This procedure, which respects the observed distribution of flux densities, spectral indices and their correlation does however not account for a possible redshift dependence of $\alpha$ and $x$. In particular, the parameter $x$ cannot be computed from a single source. As can be seen from Eq.~\eqref{eq:defx}, it is drawn from the number of sources at fixed distance $r$ from the observer, that are above the flux limit $S_*$. Since the mock samples used in~\cite{Secrest:2020has} have no notion of distance, it is not clear if their procedure, which respects the correlations between $\alpha$ and the flux density, also accounts for the correlation between $\alpha$ and $x$ \emph{as a function of distance $r$.}

In~\cite{Wang:2020ibf}, the redshift-dependence of the magnification bias $s=2x/5$ has been measured for the quasar sample of the spectroscopic survey eBOSS \citep{Dawson:2015wdb}. Between $z=1$ and $z=2$, $s$ (and therefore $x$) varies by a factor 2. There is a priory no reason why the sources used to measure the dipole in~\cite{Tiwari:2013vff,Secrest:2020has,Siewert:2020krp} would have an $x$ which is redshift-independent. If $\alpha$ also varies somehow with redshift in those samples, this could then lead to an erroneous theoretical prediction for the dipole. Even though it is unlikely that accounting for such an evolution could completely solve the tension between the observed and predicted dipole, it is worth exploring other theoretical derivations of the dipole, that would not rely on a measurement of the spectral index $\alpha$. This is what we do in the following section.

\subsection{Dipole integrated over redshift}
\label{sec:redshift}

To calculate the dipole integrated over redshift $z$, we need to compute the number of sources per solid angle $\dd \Omega$ and redshift bin $\dd z$, detected in direction $\bs{n}$, at redshift $z$, and with a flux density $S$ larger than $S_*$:
\begin{align}
\frac{\dd N}{\dd \Omega \dd z}(z, \bn, S>S_*)\, .\label{eq:dNdz}
\end{align}
This expression differs from~\eqref{eq:dNdr} by the fact that the redshift, $z$, and the redshift bin, $\dd z$, are both affected by the velocity of the observer.

To calculate the perturbations due to the observer's velocity in~\eqref{eq:dNdz}, we associate a physical distance $r(z,\bn)$ to the position $(z,\bn)$ of a source. Due to the fact that the observer is moving, sources situated on a sphere of constant redshift are at different distances $r$ from the observer in different directions. We split therefore $r(z,\bn)$ into a direction-independent part $\bar r(z)$, corresponding to the distance that an observer at rest would associate to $z$, and a perturbation $\delta r(z,\bn)$. 

We first write
\begin{align}
\frac{\dd N}{\dd \Omega \dd z}(z, \bn, S>S_*)=\frac{\dd N}{\dd \Omega \dd z}\big(r(z,\bn), \bn, L>L_*(z,\bn, \nu_s)\big)\, ,
\end{align}
and then Taylor expand this expression around $\bar r(z)$ and $\bar L_*(z, \nu_s)$ to obtain
\begin{align}
&\frac{\dd N}{\dd \Omega \dd z}\big(r(z,\bn), \bn, L>L_*(z,\bn)\big)\simeq  \label{eq:dNdzTaylor} \\
&\frac{\dd N}{\dd \bar\Omega \dd \bar z}\big(\bar r(z), L>\bar L_*(z,\nu_s)\big)\left(1-\frac{\delta\Omega}{\dd \bar\Omega}-\frac{\delta z}{1+\bar z}\right) \nonumber\\
&+\frac{\partial}{\partial r}\left(\frac{\dd \bar N}{\dd \Omega \dd z} (r,L>L_*)\right)\Big|_{r=\bar r(z)} \cdot \delta r(z,\bn)\nonumber\\
&+\frac{\partial}{\partial {L_*}}\left(\frac{\dd \bar N}{\dd \Omega \dd z} (r,L>L_*)\right)\Big|_{L_*=\bar L_*(z,\nu_s)} \cdot \delta L_*(z,\bn, \nu_s)\, .\nonumber
\end{align}
The perturbations in Eq.~\eqref{eq:dNdzTaylor} differ in three ways with respect to those in Eq.~\eqref{eq:dNdrTaylor}. First, a fixed redshift bin $\dd z$ corresponds to different physical sizes in different directions, due to the velocity of the observer. This generates an additional term, proportional to $\delta z$, in the second line of~\eqref{eq:dNdzTaylor}. Second, a fixed redshift corresponds to different distances in different directions. If the population of sources evolves with distance (i.e.\ with look-back time) this generates an additional dipolar term in the third line of~\eqref{eq:dNdzTaylor}. Finally, the perturbations in the luminosity threshold in the last line of~\eqref{eq:dNdzTaylor} differ from those in Eq.~\eqref{eq:dNdrTaylor}. In Eq.~\eqref{eq:dNdrTaylor} the luminosity perturbations are defined with respect to the background luminosity evaluated at the background frequency $\bar \nu_s$. In Eq.~\eqref{eq:dNdzTaylor} on the other hand, the frequency $\nu_s$ is the same in all directions, since $\nu_s=(1+z)\nu_o$ and we are observing at fixed $z$ and fixed $\nu_o$. As a consequence, the luminosity perturbations are defined with respect to the background luminosity evaluated at $\nu_s$. 

Let us calculate the different contributions to Eq.~\eqref{eq:dNdzTaylor} in detail. The perturbations in distance can be related to those in redshift in the following way. We split the distance into a direction-independent part plus a perturbation, both evaluated at fixed redshift $z$
\begin{align}
r(z,\bn)=\bar r(z) + \delta r(z,\bn)\, .
\end{align}
We then associate a background distance $\bar r$ and a background redshift $\bar z$ to the position $(z,\bn)$ as the ones that would be seen by an observer at rest, i.e.\ such that $\bar r(\bar z) \equiv r(z,\bn)$. We then have
\begin{align}
\bar r(z) + \delta r(z,\bn)=\bar r(\bar z)=\bar r(z-\delta z)
\simeq \bar r(z)-\frac{\partial r}{\partial z}\delta z\, .
\end{align}
From this, we obtain
\begin{align}
\delta r(z,\bn)=-\frac{\partial r}{\partial z}\delta z=\frac{1}{\HH}\bn\cdot\bv_o\,,   
\end{align}
where $\HH\equiv \dot{a}/a$ is the Hubble rate in conformal time, $a$ indicates the scale factor and a dot, a derivative with respect to conformal time $\eta$. The distance perturbation, $\delta r$, in Eq.~\eqref{eq:dNdzTaylor} is multiplied by the derivative of the number of sources with respect to distance. This term contains two contributions. The first one is due to the fact that a given pixel of size $\dd \Omega \dd z $ corresponds to a different comoving volume $\dd V$ at different distances $r$~\footnote{Note that here, we make no distinction between background and perturbed quantities, since $\delta r$ is already at first order in the velocity.}
\begin{align}
 \dd \Omega \dd z = \frac{(1+z)\HH(z)}{r^2} \dd V\, .\label{eq:Comoving_Volume}
\end{align}
And the second contribution is due to the fact that the number of sources in a comoving volume $\dd V$ does generally evolve as a function of distance, i.e. as a function of look-back time. We obtain
\begin{align}
\frac{\partial}{\partial r}\left(\frac{\dd \bar N}{\dd \Omega \dd z} (r,L>L_*)\right)\Big|_{r=\bar r} & =\HH\left(-1+\frac{2}{\bar r\HH} +\frac{\dot\HH}{\HH^2}-f_{\rm evol} \right)\nonumber\\
&~~ \times \frac{\dd \bar N}{\dd \Omega \dd z} (\bar{r},L>L_*) \, , \label{eq:ddr_ln_dNdOmegadz}
\end{align}
where the last term, $f_{\rm evol}$, encodes the evolution of sources with time
\begin{align}
f_{\rm evol} & \equiv \frac{1}{\HH}\left(\frac{\dd \bar N}{\dd V}(r,L>L_*)\right)^{-1}\frac{\partial}{\partial\eta}\left(\frac{\dd \bar N}{\dd V}(r,L>L_*)\right) \nonumber\\
& = -\frac{\p \ln \left(\frac{\dd \bar N}{\dd V}(z,L>L_*)\right)}{\p \ln (1+z)}\, .\label{eq:def_fevol}
\end{align}
Here again we can neglect the difference between $f_{\rm evol}(r)$ and $f_{\rm evol}(z)$ since it would lead to a correction to the dipole quadratic in the velocity.

The last line in Eq.~\eqref{eq:dNdzTaylor} contains the perturbations in the luminosity threshold that can be calculated using that 
\begin{align}
L(z,\bn,\nu_s)=4\pi\frac{d^2_L(z,\bn)}{1+z}S(\nu_o)    
\end{align}
and
\begin{align}
\bar L(z,\nu_s)=4\pi\frac{\bar d^2_L(z)}{1+z}S(\nu_o)    \label{eq:dLback}
\end{align}
leading to
\begin{align}
\delta L_* (z,\bn,\nu_s)   =\bar L_*(z,\nu_s) \frac{2\delta d_L(z,\bn)}{\bar d_L(z)}\,  . \label{eq:dLz}
\end{align}
Comparing with Eq.~\eqref{eq:dLr}, we see that the perturbations in the luminosity threshold differ. First, here the luminosity threshold does not depend on the spectral index of the sources $\alpha$. This is due to the fact that since we observe sources at the same redshift in all directions, we automatically access the same part of the spectrum, namely the frequency $\nu_s=(1+z)\nu_o$. As such the fact that the spectrum varies with $\nu_s$ has no impact on the luminosity threshold. In the previous calculation on the other hand, we were observing sources at the same distance $r$ in all directions: these sources were located at different redshifts due to the observer's velocity, meaning that we were not observing the same part of the spectrum in different directions. The second difference between Eqs.~\eqref{eq:dLr} and~\eqref{eq:dLz} is that the perturbations in the luminosity distance at fixed distance $r$ are not the same as the ones at fixed redshift $z$. In particular, we have~\citep{Bonvin:2005ps}
\begin{align}
\frac{\delta d_L(z,\bn)}{\bar d_L(z)}=\frac{1}{\HH \bar r}\bn\cdot\bv_o\, , \label{eq:dLpertz} 
\end{align}
which differs from Eq.~\eqref{eq:deltadLr}. The perturbations in the luminosity threshold in Eq.~\eqref{eq:dNdzTaylor} are multiplied by the derivative of the number of sources with luminosity, which is given in~\eqref{eq:xterm} and depends as before on the magnification bias parameter $x$.

Combining everything and integrating over redshift we find for the dipole amplitude
\begin{align}
\mathcal{D}^z_{\rm kin}=\int_0^\infty \dd z f(z) 
\left[2+\frac{2(1-x)}{\bar r\HH} +\frac{\dot\HH}{\HH^2}-f_{\rm evol} \right] |\bv_o|\,.\label{eq:Dz}
\end{align}
Here, $f(z)=f(\bar z)=f(r)/(\HH(1+z))$ since the difference between the distribution of sources in real space and in redshift space due to the observer's velocity would lead to an effect at second order in the velocity.  Eq.~\eqref{eq:Dz} is consistent with the results derived in~\cite{Maartens:2017qoa} for the galaxy number counts at fixed redshift.

Note that the term proportional to $(\mathcal{H}\bar{r})^{-1}$ in Eq.~\eqref{eq:Dz} diverges if the number density does not vanish for redshift $z\to 0$. This can be the case if $f(z)$ represents the observed redshift distribution of sources instead of the background redshift distribution. Indeed, sources very close to the observer can have a negative Doppler shift (due to their peculiar velocity towards the observer) which is larger than their background redshift. This divergence, however, is an artifact of considering too low redshift sources, for which higher order corrections would be needed. Instead, a redshift cutoff $z\e{\epsilon}>0$ on the lower boundary of the integral in \eqref{eq:Dz} may be implemented or negative redshift sources may be ignored when creating the interpolation function, by forcing $f(z=0)=0$.

\section{Comparison}

Eqs.~\eqref{eq:Dchi} and~\eqref{eq:Dz} must be mathematically equivalent, since from an observational point of view, summing over all sources does not require knowledge of their redshift or distance. The equivalence between the two expressions can easily be demonstrated starting from Eq.~\eqref{eq:dNdzTaylor} and rewriting (see also Appendix A in~\cite{Nadolny:2021hti} for a similar derivation)
\begin{align}
\frac{\partial}{\partial r}\Bigg(\frac{\dd \bar N}{\dd \Omega \dd z} (r,&L>L_*)\Bigg)=\HH(1+z) \Bigg\{
\frac{\dd}{\dd z}\left(\frac{\dd \bar N}{\dd \Omega \dd z} (r,L>L_*)\right)\nonumber\\
&-\frac{\partial}{\partial L_*}\left(\frac{\dd \bar N}{\dd \Omega \dd z} (r,L>L_*)\right)\frac{\dd L_*}{\dd z}\Bigg\}\, .
\end{align}

From Eq.~\eqref{eq:dLback} we can relate the luminosity density to the flux density at fixed frequency $\nu_s$
\begin{align}
\bar L_*(z,\nu_s)=4\pi (1+z)\bar r^2(z)S_*(\nu_o)=4\pi (1+z)^{1+\alpha}\bar r^2(z)S_*(\nu_s)\, ,   
\end{align}
where we have used that $S_*(\nu_o)\propto \nu_o^{-\alpha}\propto \nu_s^{-\alpha}(1+z)^\alpha$. With this we have
\begin{align}
\frac{\dd L_*}{\dd z}=\left(1+\alpha+\frac{2}{\bar r\HH}\right)\frac{L_*}{1+z} \, .  \label{eq:dLdz}
\end{align}
Eq.~\eqref{eq:dNdzTaylor} can then be rewritten as
\begin{align}
&\frac{\dd N}{\dd \Omega \dd z}\big(r(z,\bn), \bn, L>L_*(z,\bn)\big)= \label{eq:comp}\\
&\frac{\dd N}{\dd \bar\Omega \dd \bar z}\big(\bar r, L>\bar L_*(z,\nu_s)\big)\left(1+3\bn\cdot\bv_o\right)\nonumber\\
&+(1+z)\frac{\dd}{\dd z}\left(\frac{\dd \bar N}{\dd \Omega \dd z} (r,L>L_*)\right) \bn\cdot \bv_o\nonumber\\
&+\frac{\dd \bar N}{\dd \Omega \dd z} (r,L>L_*)x(1+\alpha)\bn\cdot\bv_o\, ,\nonumber
\end{align}
where the $2/(\bar{r}\HH)$ term in~\eqref{eq:dLdz} cancels the one in~\eqref{eq:dLpertz}. Integrating  Eq.~\eqref{eq:comp} over redshift, and using the fact that the term in the third line can be integrated by parts\footnote{This step requires that boundary terms vanish.}, we obtain, after subtracting the angular average, Eq.~\eqref{eq:Dchi}. As already pointed out in~\cite{Nadolny:2021hti}, the relativistic expression for the galaxy number counts is therefore consistent with the expression of Ellis and Baldwin in Eq.~\eqref{eq:dipEB}, once averaged over all sources, provided that $\alpha$ and $x$ are redshift independent.

\label{sec:comp}
\begin{figure}
    \centering
    \includegraphics[width=0.46\textwidth]{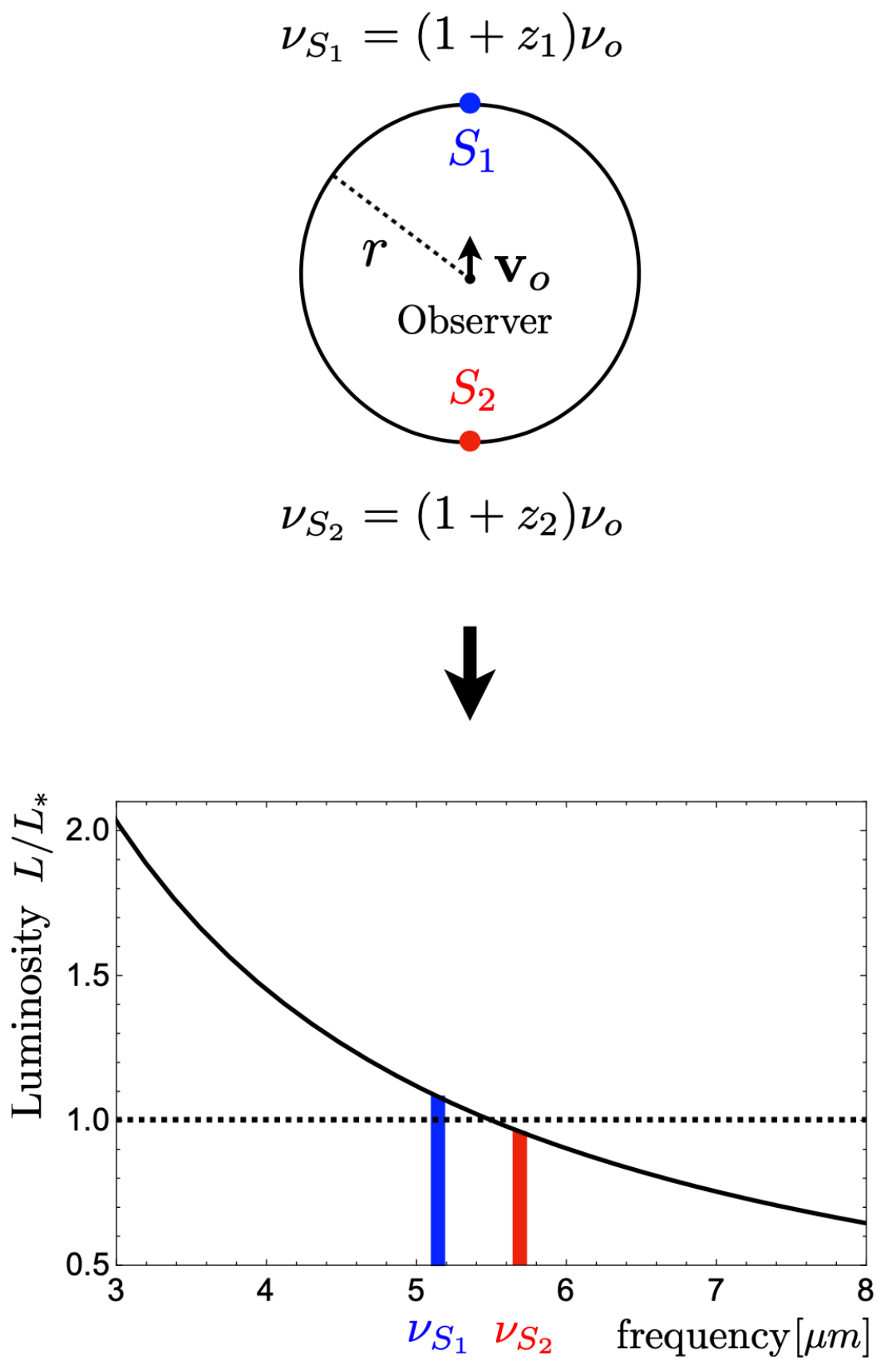}
    \caption{
    At fixed distance $r$ from the observer, the source frequency $\nu_{S_1}$    (associated to the observed frequency $\nu_o$) is smaller than $\nu_{S_2}$ for an observer moving in the direction of $S_1$. 
    Since $L\propto \nu_s^{-\alpha}$ (with $\alpha>0$) the observer accesses a part of the spectrum of $S_1$ which is above the luminosity threshold $L_*$, whereas for $S_2$ the luminosity is below the luminosity threshold.
    More sources are therefore observed in the direction of the observer's velocity, which generates an additional contribution in the dipole. }
    \label{fig:Dr}
\end{figure}

\begin{figure}
    \centering
    \includegraphics[width=0.35\textwidth]{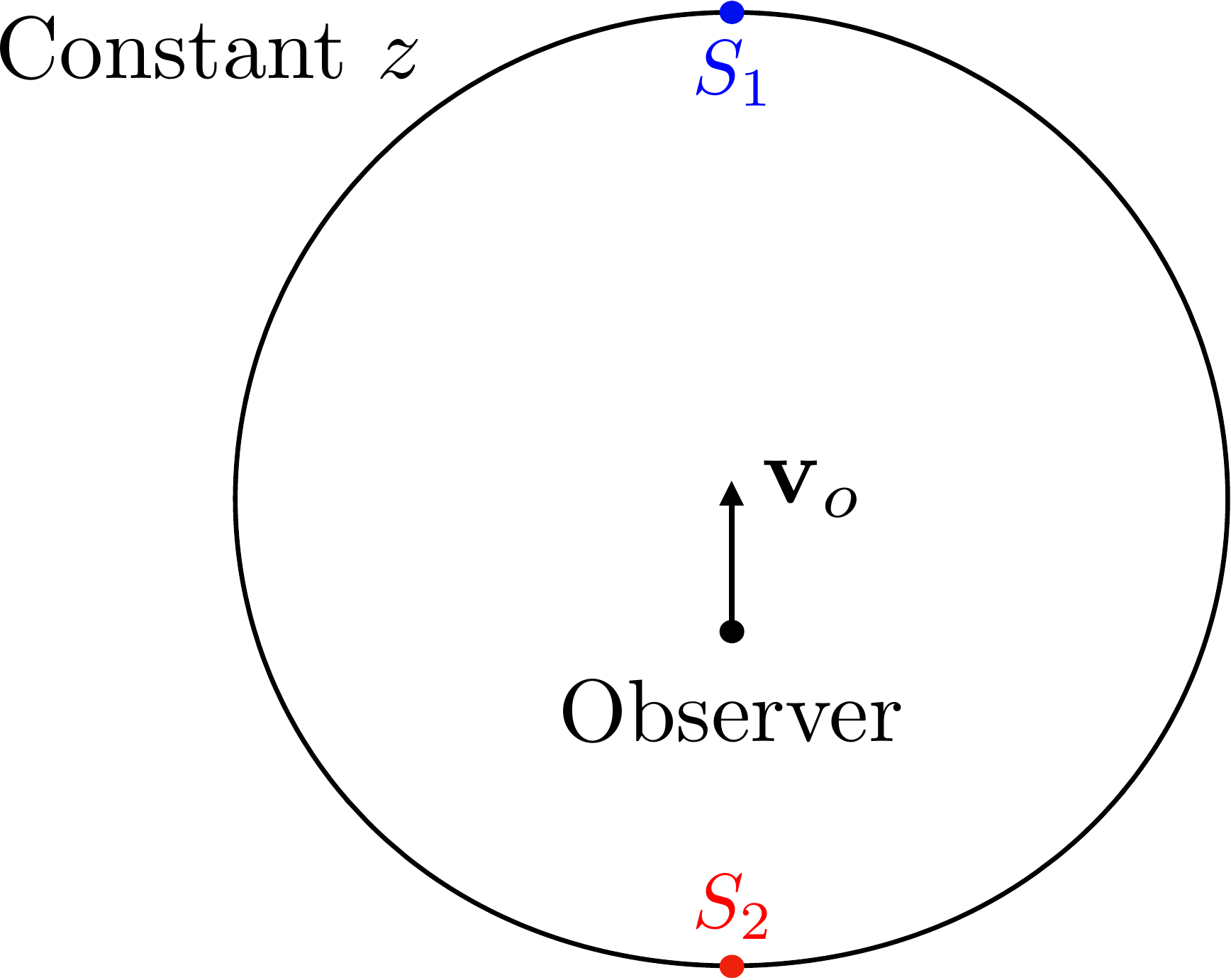}
    \caption{Sources at constant redshift from a moving observer are located at different physical distances depending on the direction of observation. 
    }
    \label{fig:Dz}
\end{figure}

Even though Eqs.~\eqref{eq:Dchi} and~\eqref{eq:Dz} are mathematically equivalent, it is interesting to use these two different theoretical expressions to predict the amplitude of the dipole, since they depend on different parameters. Eq.~\eqref{eq:Dchi} depends indeed on the parameter $\alpha$ which describes the shape of the source spectra, whereas Eq.~\eqref{eq:Dz} depends on the parameter $f_{\rm evol}$ which describes the evolution of the population of sources with time. These different dependencies on the properties of the sources can be easily understood by looking at Figs.~\ref{fig:Dr} and~\ref{fig:Dz}. When integrating over distance, one compares the number of sources in different directions at fixed distance from the observer. A source located in the direction of the observer's velocity, $S_1$, will have a smaller redshift than a source located in the opposite direction, $S_2$, see Fig.~\ref{fig:Dr}. As a consequence, the observed frequency $\nu_o$ which is the same in both direction, corresponds to a source frequency $\nu_{S_1}$ which is smaller than $\nu_{S_2}$. As can be seen from Fig.~\ref{fig:Dr}, this means that the luminosity density of the source $S_1$ is larger than the one of the source $S_2$. More sources above the threshold $L_*$ can therefore be detected in the direction of the observer's velocity.

The redshift-dependent result in Eq.~\eqref{eq:Dz} is not affected by this effect: two sources in opposite directions have the same redshift, and therefore the same $\nu_s$. As a consequence, the same part of the spectrum is observed in both directions, corresponding to the same luminosity density. Another effect contributes however in this case: as can be seen from Fig.~\ref{fig:Dz}, two sources in opposite directions at the same redshift are located at different distances from the observer. The source aligned with the observer's velocity, $S_1$, is further away from the observer than the source in the opposite direction, $S_2$. If the number of sources per comoving volume evolves with distance (i.e. with look-back time) then the two observers will see a different number of sources. This effect is encoded in the parameter $f_{\rm evol}$. It depends on the intrinsic evolution of the population of sources but also on selection effects.

Finally, let us mention that the dipole in the number counts of optical galaxies is not affected by the effect depicted in Fig.~\ref{fig:Dr}. The spectra of these galaxies have usually a continuum which is indeed relatively flat within the observed frequency range~\citep[see e.g.][]{1992ApJ...388..310K}, leading to $\frac{\partial }{\partial \nu_s}\bar L_*(r,\nu_s)=0$ in Eq.~\eqref{eq:Lalpha}. As a consequence, even though the frequency range is slightly different in different directions, due to the observer's velocity, the total luminosity (integrated over the frequency range) remains the same. The term proportional to $\alpha$ in Eq.~\eqref{eq:Dchi} vanishes therefore for optical galaxies. On the other hand, the effect depicted in Fig.~\ref{fig:Dz} is present for all sources that are evolving with redshift.

\section{Application to concrete examples}
\label{sec:example}
We now calculate the dipole expected in two different samples of quasars, using the integral over redshift given in Eq.~\eqref{eq:Dz}. We first briefly discuss the expected random contaminations that arise when comparing the predicted dipole with observations. We then apply our findings to the eBOSS Quasar sample (QSO) for which the parameters $x(z)$ and $f_{\rm evol}(z)$ have been measured~\citep{Wang:2020ibf}.
Finally, we analyze the expected kinematic dipole for the CatWISE2020 catalog~\citep{Secrest:2020has}, using different assumptions for $x(z)$ and $f_{\rm evol}(z)$.

\subsection{Signal and contaminations}

The signal, i.e.\ the kinematic dipole, can be split into three parts 
\begin{align}
\mathcal{D}\e{kin}= \mathcal{D}\e{cosmo}+ \mathcal{D}\e{mag} + \mathcal{D}\e{evol} \,.
\end{align}
Note that here, we drop the superscript $z$ in the kinematic dipole since we always use the integral over redshift~\eqref{eq:Dz}. The cosmological part depends only on the background evolution of the Universe, but not on the properties of the sources apart from their redshift distribution $f(z)$
\begin{align}
\mathcal{D}_{\rm cosmo}\equiv \int_0^\infty \dd z f(z) 
\left[2+\frac{2}{\bar r\HH} +\frac{\dot\HH}{\HH^2} \right]|\bv_o|\, .  \label{eq:D_Cosmo}
\end{align}
The magnification part encodes the magnification of the flux due to the observer's velocity, and depends therefore on the function $x(z)$
\begin{align}\label{eq:Dmag}
\mathcal{D}_{\rm mag}\equiv-\int_0^\infty \dd z f(z) 
\frac{2 x(z)}{\bar r\HH} \, |\bv_o|\,.
\end{align}
The evolution part is due to the evolution of the source population with time
\begin{align}\label{eq:Devol_integral}
\mathcal{D}_{\rm evol}\equiv-\int_0^\infty \dd z f(z) 
f_{\rm evol}(z)\, |\bv_o|\,.
\end{align}

In addition to these kinematic contributions, there are two contaminations to the observed dipole: a contribution from intrinsic clustering and a contribution from shot noise. These contributions are described in detail in~\cite{Nadolny:2021hti}, here we simply summarize their results. We associate a vector to the kinematic dipole
\be
\bs{\mathcal{D}}_{\rm kin}\equiv \mathcal{D}_{\rm kin}\cdot \hat{\bv}_o\, ,
\ee
which points in the direction of the observer's velocity. Intrinsic clustering and shot noise have dipolar contributions which are a priori not aligned with the kinematic one. The total dipole vector is the sum of the three contributions
\begin{align}
\bs{\mathcal{D}}=\bs{\mathcal{D}}\e{kin}+ \bs{\mathcal{D}}\e{int} + \bs{\mathcal{D}}\e{sn} \ \in \mathbb{R}^3\,.
\end{align}
The intrinsic contribution comes from the fact that the distribution of sources is not perfectly homogeneous and isotropic in our Universe, even on very large scales. Hence we expect an intrinsic dipolar modulation in the source number counts. The direction and amplitude of this intrinsic dipole are unknown (they depend indeed on the initial conditions). Since we know the statistical properties of the initial conditions, we can however calculate the variance of this contribution (see~\cite{Secrest:2020has, Nadolny:2021hti}): $\langle\mathcal{D}\e{int}^2\rangle=3\sigma_{\rm int}^2=9C_1^{\rm int}/(4\pi) $, where $C_1^{\rm int}$ is the dipole of the clustering angular power spectrum, which can be calculated for example with CLASS~\citep{Blas:2011rf}. The shot noise contribution, $\mathcal{D}\e{sn}$, comes from probing a smooth field with a discrete sample. Its direction and amplitude are also random. The variance of this contribution scales as $\langle\mathcal{D}\e{sn}^2\rangle =3\sigma_{\rm sn}^2 =9 f_{\rm sky} /N$, where $N$ is the total number of sources used to measure the dipole and $f_{\rm sky}$ is the observed fraction of the sky. Note that if the signal is expanded on spherical harmonics, having an incomplete sky coverage leads to a leakage from higher multipoles to the dipole, which roughly increases the shot noise contribution by a factor $f_{\rm sky}^{-1}$~\citep{Nadolny:2021hti}. It is however possible to avoid such a leakage by explicitly seeking a dipolar pattern, instead of expanding on spherical harmonics and extract the dipole from this expansion. This is for example the method used in~\cite{Secrest:2020has}. In this case there is no contamination from higher multipoles, as long as those are significantly smaller than the dipole (which is the case here, since the higher multipoles are not affected by the observer velocity at linear order). 

Following~\cite{Nadolny:2021hti}, we can compute the expectation value of the theoretical dipole amplitude. We choose the $z$-direction to be aligned with the kinematic dipole vector. The individual components of the theoretical dipole $\bs{\mathcal{D}}=(\mathcal{D}^x, \mathcal{D}^y, \mathcal{D}^z)$ follow a Gaussian distribution with variance
$\sigma_r^2 = \sigma\e{int}^2 +\sigma\e{sn}^2$ and expectation value $\langle \mathcal{D}^x\rangle =\langle \mathcal{D}^y\rangle =0 $, and $\langle \mathcal{D}^z\rangle = \mathcal{D}\e{kin}$. Therefore, the amplitude of $\bs{\mathcal{D}}$ follows a non-central $\chi(3)$-distribution, whose expectation value and variance may be expressed with the generalized Laguerre polynomial
\begin{align}
\langle \mathcal{D}\rangle =&  \sigma_r \sqrt{\frac{\pi}{ 2}} L_{1/2}^{1/2}\left( - \frac{\mathcal{D}\e{kin}^2}{2 \sigma_r^2} \right)\,, \label{eq:D_Theory} \\
\sigma_{\mathcal{D}}^2 = & 3 \sigma_r^2 +\mathcal{D}\e{kin}^2 - \langle \mathcal{D} \rangle^2  \,. \label{eq:sigma_D_Theory}
\end{align}
The tension between the observed and theoretical dipole amplitudes is then given by
\begin{align}
T = \frac{| \langle \mathcal{D}\rangle - \mathcal{D}\e{obs}|}{\sigma_{\mathcal{D}}} \sigma \,. 
\label{eq:tension}
\end{align}
In the next subsections, we compute the expectation value of $ \mathcal{D}$ and its corresponding variance for different source catalogues.

\subsection{The eBOSS Quasar sample}

We first calculate the dipole expected from the Quasar (QSO) sample of eBOSS~\citep{Ross:2020lqz}. Since the sky coverage of this sample is relatively small (4'808 deg$^2$) and the number of sources is of 343'708 only, this sample is not optimal to measure a dipolar modulation. However the magnification bias $s(z)=x(z)/2.5$ and the evolution bias $b_e(z)\equiv f_{\rm evol}(z)$ have been measured recently for a sub-sample of 13'876 quasars~\citep{Wang:2020ibf}, allowing a precise theoretical prediction of the dipole for this catalog. It is therefore instructive to study this sample as a starting point.

\begin{figure}
\centering
\includegraphics[width=0.45\textwidth]{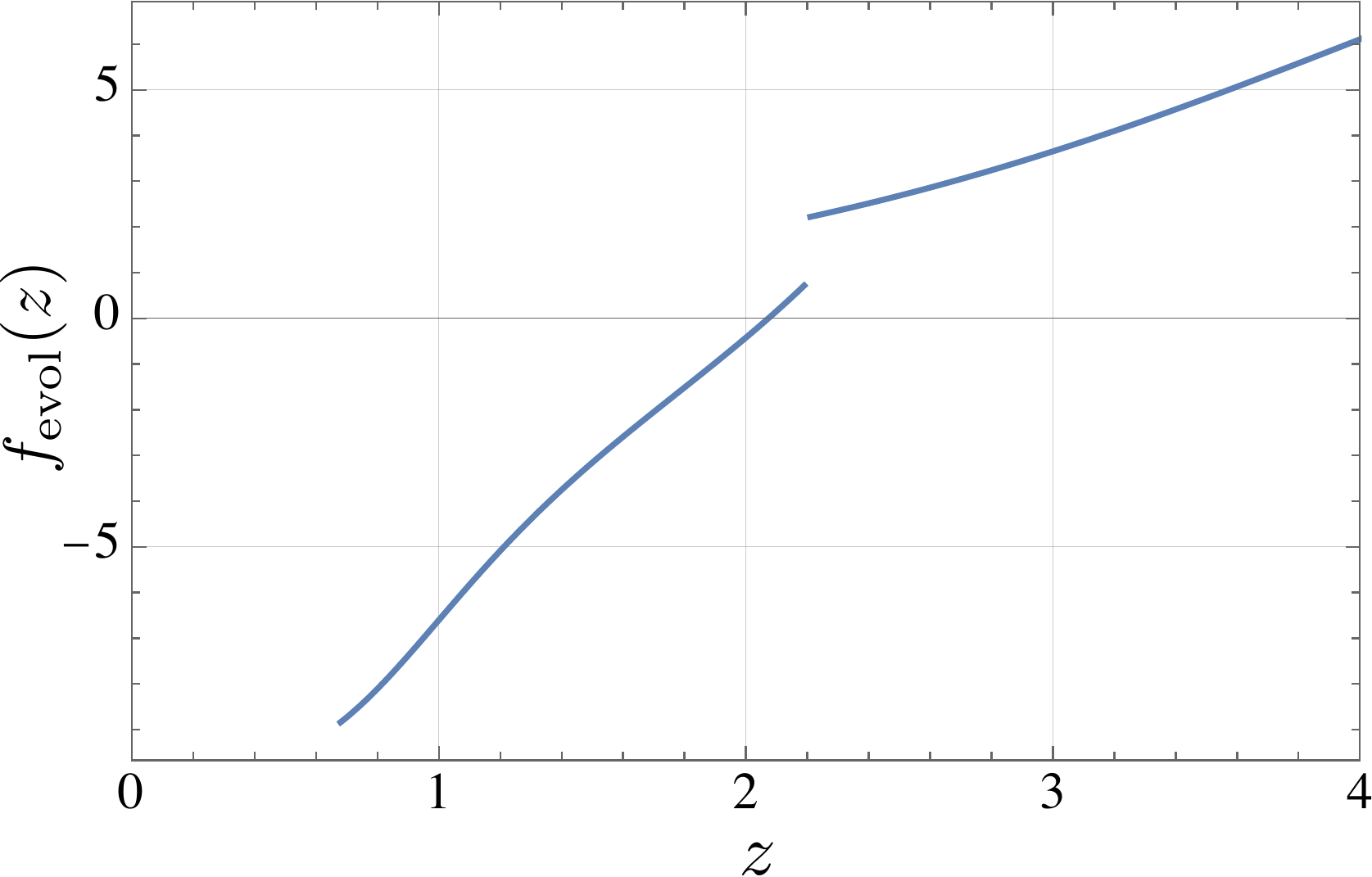}
\caption{Evolution bias as a function of redshift for the eBOSS Quasar sample computed using the bestfit parameters of \citet{Wang:2020ibf} in Eq.~\eqref{eq:luminosity_function}. 
The discontinuity at $z=z_{\rm p}$ is ill understood and may be due to unknown systematics.
}
\label{fig:fevol}
\end{figure}

We start by summarizing the results of~\cite{Wang:2020ibf} and then use them to calculate the dipole. The comoving number density of quasars can be expressed as an integral of the luminosity function $\phi(M,z)$ over absolute magnitudes below some threshold $M_*$
\begin{align}
\frac{\dd \bar{N} }{\dd V}(z,M< M_*) = \int_{-\infty}^{M_*} \dd M \phi(M,z)\,. \label{eq:Comoving_Number_Density}
\end{align}
The luminosity function can be parametrized in the following way
\begin{align}\label{eq:luminosity_function}
\phi(M,z) & = \frac{\phi_*}{10^{0.4(\hat\alpha + 1)(M-\hat M(z))} + 10^{0.4 (\hat\beta +1)(M-\hat M(z))}}\,,
\end{align}
with
\begin{align}
\hat M(z) & = \hat M(z\e{p}) - \frac{5}{2} \l[k_1(z-z\e{p}) + k_2(z-z\e{p})^2 \r]\,.
\end{align}
Here $z\e{p}=2.2$ denotes a pivot redshift and $\phi_*, \hat M(z_{\rm p})$, $\hat\alpha$, $\hat\beta$, $k_1$ and $k_2$ are parameters that can be fitted to the data. Note that $\hat\alpha$, $\hat\beta$, $k_1$ and $k_2$ can take different values for redshift below or above $z\e{p}$. We use the best fit values of \cite{Wang:2020ibf}, given in their~\footnote{Note that the values for $\phi_*$ and $\hat M(z_{\rm p})$ have accidentally been swapped in their table. The correct values are $\phi_* = 10^{-5.76}$ and $\hat M(z_{\rm p}) = -26.20$.} Table\,1 and compute $f_{\rm evol}$ using Eq.\,\eqref{eq:def_fevol}. We see that for redshift 1, $f_{\rm evol}\sim -7$ and flips sign around redshift 2, as can be seen on Fig.\,\ref{fig:fevol}. This can be understood by the formation of quasars before redshift 2 and their more efficient disappearance after that. The end of individual quasar activity and merging are two reasons why $f\e{evol}$ could be negative. This translates to a change in the sign of $\mathcal{D}\e{evol}$ if sources are more abundant below or above redshift $2$.

\begin{figure}
\centering
\includegraphics[width=0.45\textwidth]{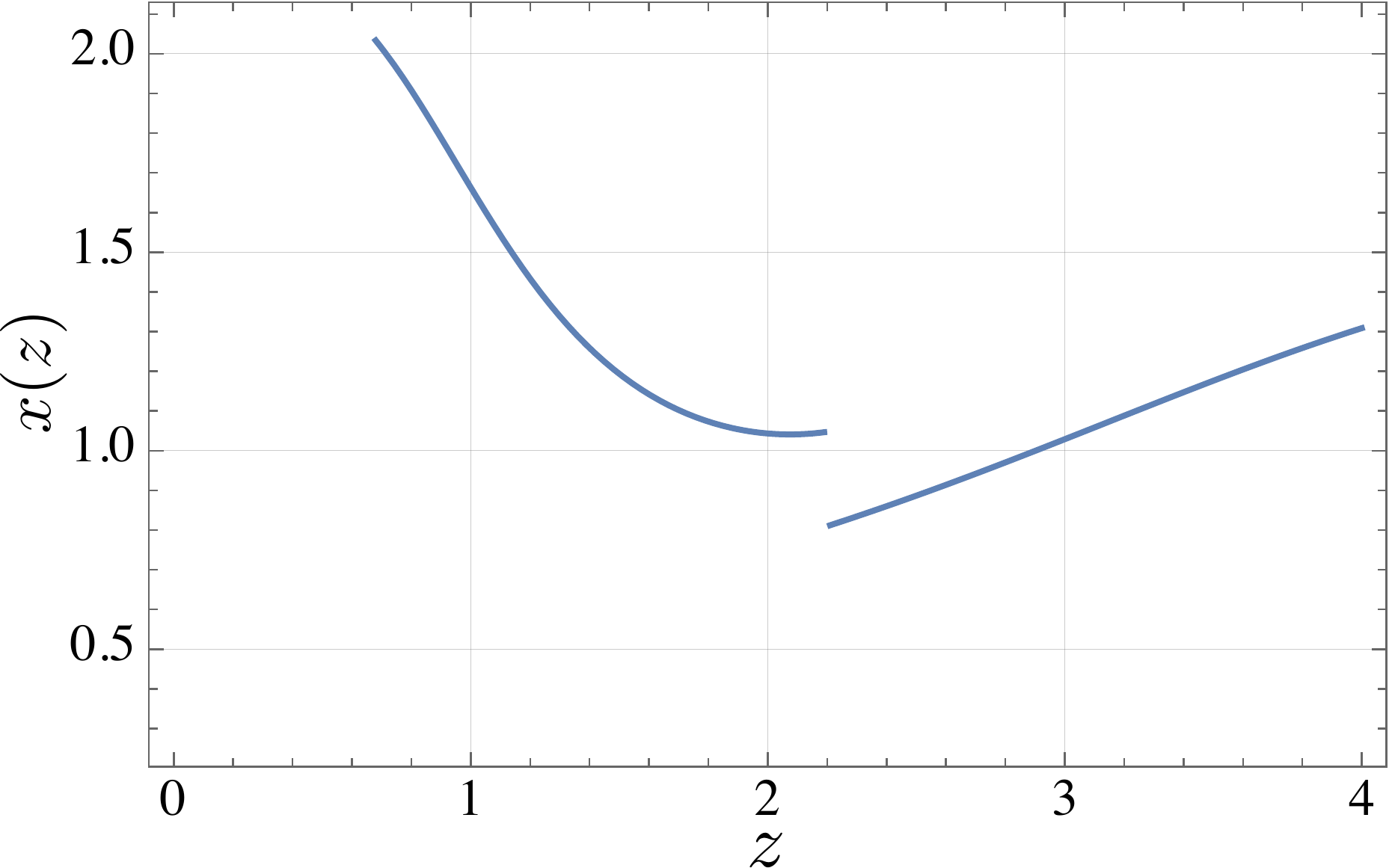}
\caption{The parameter $x(z)= \frac{5}{2}s(z)$ as a function of redshift for the eBOSS Quasar sample computed using the bestfit parameters of \citet{Wang:2020ibf} in Eq.~\eqref{eq:luminosity_function}. }
\label{fig:magnification_bias}
\end{figure}
\begin{figure}
\centering
\includegraphics[width=0.45\textwidth]{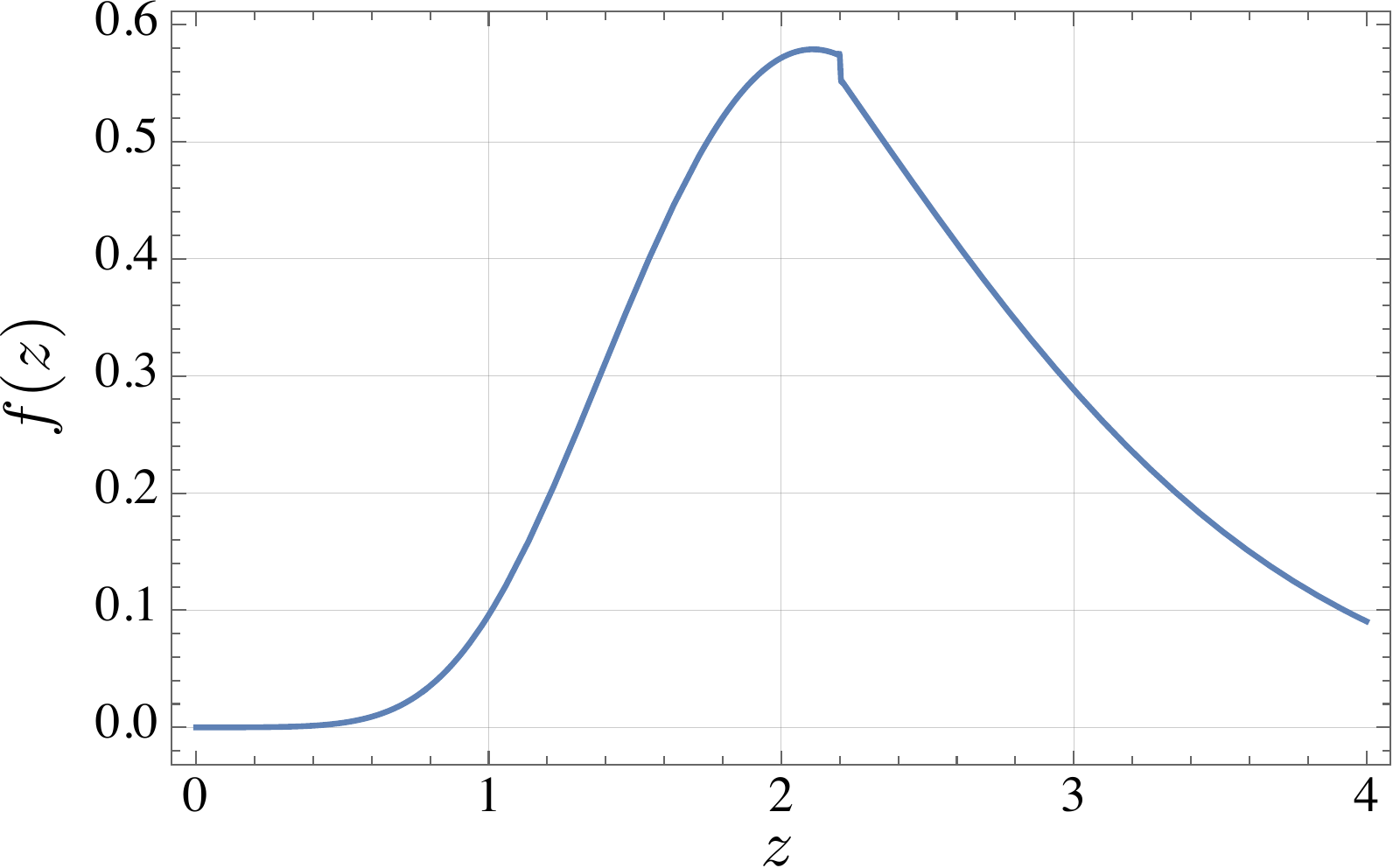}
\caption{We plot the redshift distribution of sources for the eBOSS Quasar sample computed using Eqs.\,\eqref{eq:Comoving_Number_Density} and \eqref{eq:Comoving_Volume}. The redshift distribution is peaked at redshift $2$ and averages to $\langle z \rangle = 2.28$.}
\label{fig:QSO_Redshift_Distribtion}
\end{figure}

In Fig.~\ref{fig:magnification_bias} we plot the magnification bias $x(z)$. Note that there is a factor $h^3$ missing in Fig.\,3 of~\cite{Wang:2020ibf}, which is the result of working with different units for $\phi$ and $\dd N/ \dd V$. 
The discontinuities in both $x(z)$ and $f\e{evol}(z)$ at $z_{\rm p}$ is ill understood and could be attributed to unknown systematics \citep{Kulkarni2019}. We also plot the normalized redshift distribution of sources in Fig.~\ref{fig:QSO_Redshift_Distribtion}. 

We can now compute $\mathcal{D}\e{cosmo}$ in Eq.~\eqref{eq:D_Cosmo} using the redshift distribution plotted in Fig.~\ref{fig:QSO_Redshift_Distribtion}. Assuming a $\Lambda$CDM Universe with best fit parameters from Planck~\citep{Aghanim:2018eyx} and using the observer's velocity measured from the CMB we find
\begin{align}
\mathcal{D}\e{cosmo}= 0.0040\,.
\label{eq:DcosmoeBOSS}
\end{align}
The magnification dipole in Eq.~\eqref{eq:Dmag} can be calculated using $x(z)$ plotted in Fig.\,\ref{fig:magnification_bias}. We obtain
\begin{align}
\mathcal{D}\e{mag}=-0.0022 \,. 
\end{align}
This contribution is always negative since $x(z)$ is always positive: the cumulative number of sources can indeed only decrease with increasing $S_*$. This therefore decreases the total amplitude of the kinematic dipole. Let us mention that the effective magnification bias averaged over the redshift distribution is 
$x\e{eff}=1.1$, which is close to the standard assumption of $x\e{eff}\simeq 1$ often used in the literature, but
smaller than the $x\e{eff}=1.7$ of the CatWISE2020 sample~\citep{Secrest:2020has}.

The evolution dipole $\mathcal{D}\e{evol}$ in Eq.~\eqref{eq:Devol_integral} is obtained using $f_{\rm evol}$ plotted in Fig.~\ref{fig:fevol} and we find
\begin{align}
\mathcal{D}\e{evol} = -0.0007\,.
\end{align}
The evolution dipole is almost $6$ times smaller than the cosmological part and it is negative. This is due to the fact that the source distribution is peaked around redshift $2$, where the evolution bias $f_{\rm evol}$ changes sign. The contributions below and above redshift $2$ therefore partially cancel out, resulting in a small negative contribution. From Fig.~\ref{fig:magnification_bias}, we can immediately guess
that for a source distribution that would be peaked at lower redshift, where the number of quasars rapidly decreases, the evolution dipole would be significantly larger and positive. Summing the three contributions we predict that the observed quasar count kinematical dipole should be 
\begin{align}
\mathcal{D}\e{kin}= \mathcal{D}\e{cosmo}+ \mathcal{D}\e{mag} + \mathcal{D}\e{evol} = 0.0011
\,.\label{eq:DkineBOSS}
\end{align}
The intrinsic dipole contamination can be estimated using CLASS \citep{Blas:2011rf}. Assuming that the bias of the eBOSS quasars evolves as $b(z)=1.2/D(z)$~\citep{Wang:2020ibf}, where $D(z)$ is the linear growth rate we obtain $\sqrt{\langle\mathcal{D}\e{int}^2\rangle} = 0.0003$, which is about a third of the kinematic dipole. 
The subsample of quasars used to measure the evolution bias and magnification bias contains only 13'876 sources. However one can assume that these quantities are representative of the full sample and use all 343'708 quasars to measure the dipole. In this case the shot noise contamination is $\sqrt{\langle\mathcal{D}\e{sn}^2\rangle} = 0.0018$. This is larger than the expected signal in Eq.~\eqref{eq:DkineBOSS}. Moreover, since the sky coverage is relatively small: $f_{\rm sky}=0.12$, it may be challenging to find a dipolar pattern in the data. This indicates that the eBOSS quasars sample is probably too small to allow for a robust detection of the kinematic dipole. 

\subsection{CatWISE2020}

\begin{figure}
    \centering
    \includegraphics[width=0.48\textwidth]{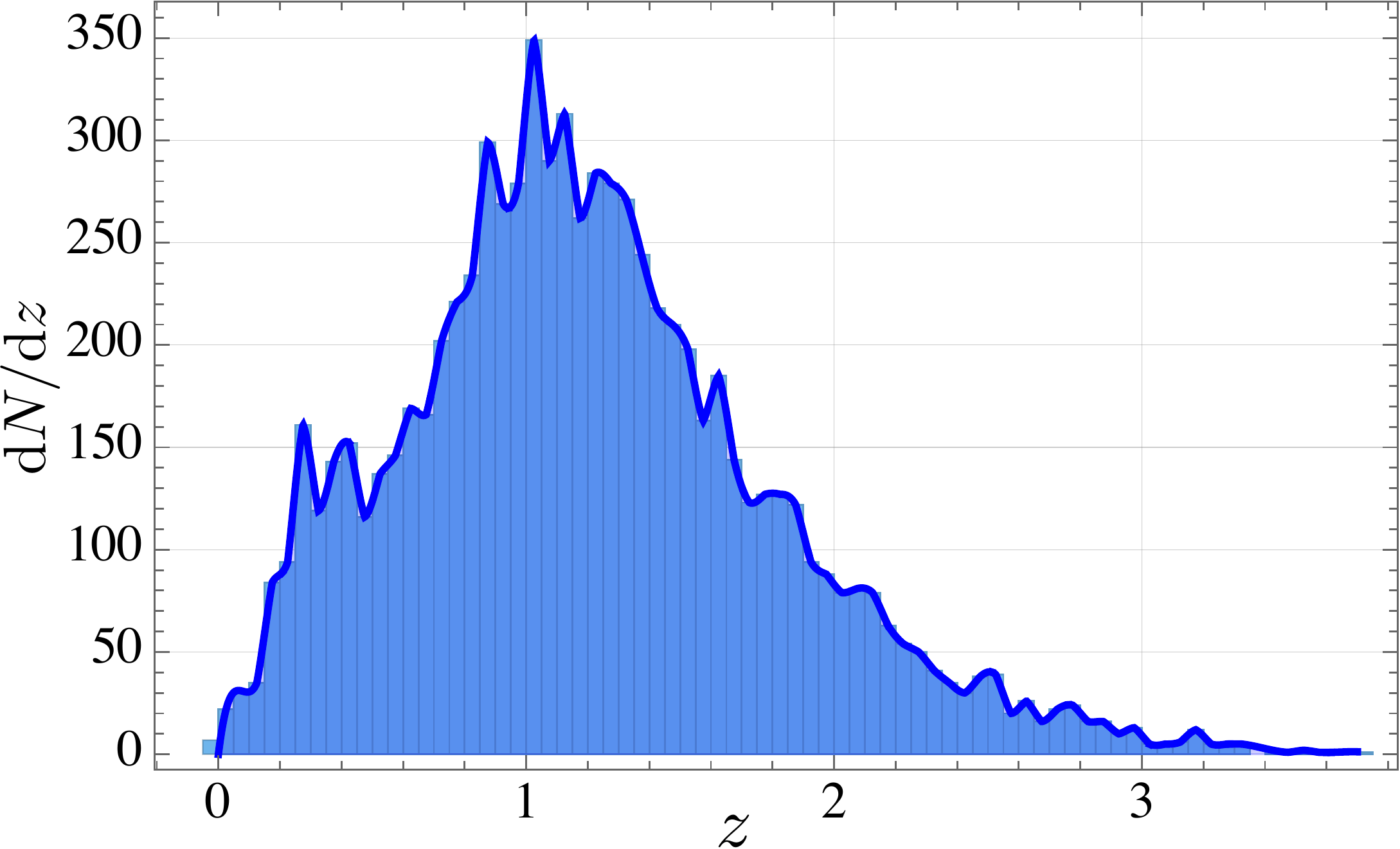}
    \caption{Unnormalized redshift distribution, within 75 redshift bins which span $z\in [0,3.7]$. A cubic interpolation is plotted, imposing $f(z\to 0)\to 0$, which gives similar results for $\mathcal{D}\e{kin}$ as imposing a lower redshift cutoff and avoid unphysical divergences coming from $\mathcal{D}\e{cosmo}$ and $\mathcal{D}\e{mag}$.}
    \label{fig:dNoverdz}
\end{figure}

We now use the information gained with the eBOSS sample to discuss the expected signal in the CatWISE2020 sample. The dipole from this catalog has been observed to be $|\bs{\mathcal{D}}\e{obs}|= 0.0155$, i.e.\ nearly twice as large as the prediction from Eq.~\eqref{eq:dipEB}, $\mathcal{D}\e{EB}= 0.0072$,~\citep{Secrest:2020has}, and pointing in a direction which is $27.8$° away from the CMB dipole. The distribution function $f(z)$ has been estimated by cross-matching a subsample of the sources with the SDSS survey, which provides spectroscopic redshift information. This distribution is shown in Fig.\,\ref{fig:dNoverdz}, together with the unnormalized cubic interpolation that we use in our theoretical prediction. In contrast to the previous sample, this distribution peaks at redshift $1$ and does contain negative observed redshift sources. This implies that in principle, the number density does not vanish as $z\to 0$, which could lead to divergences in Eqs.~\eqref{eq:D_Cosmo} and \eqref{eq:Dmag}. However, those divergences are an artifact of not computing higher order corrections to the dipole: a negative redshift means indeed that the Doppler contribution from the observer and source peculiar velocities is larger than the background contribution. In this case, perturbation theory breaks down, and the linear prediction in Eq.~\eqref{eq:Dz} is not valid anymore. We can remove the divergences by either using a lower redshift cutoff or by forcing the interpolation to be $f(z=0)=0$. We chose the latter and checked that those two approaches give similar results for $\mathcal{D}\e{kin}$.

As before, the cosmological part of the dipole can be calculated in $\Lambda$CDM using the observer's velocity measured from the CMB. We find 
\be
\mathcal{D}_{\rm cosmo}=0.0066\, .
\ee
Comparing with Eq.~\eqref{eq:DcosmoeBOSS} we see that the cosmological part is larger for the CatWISE2020 sample than for the eBOSS sample. This is due to the fact that the CatWISE2020 sample contains more sources at lower redshift, for which the $1/(\bar r\HH)$ in Eq.~\eqref{eq:D_Cosmo} is larger.
\begin{figure}
    \centering
    \includegraphics[width=0.47\textwidth]{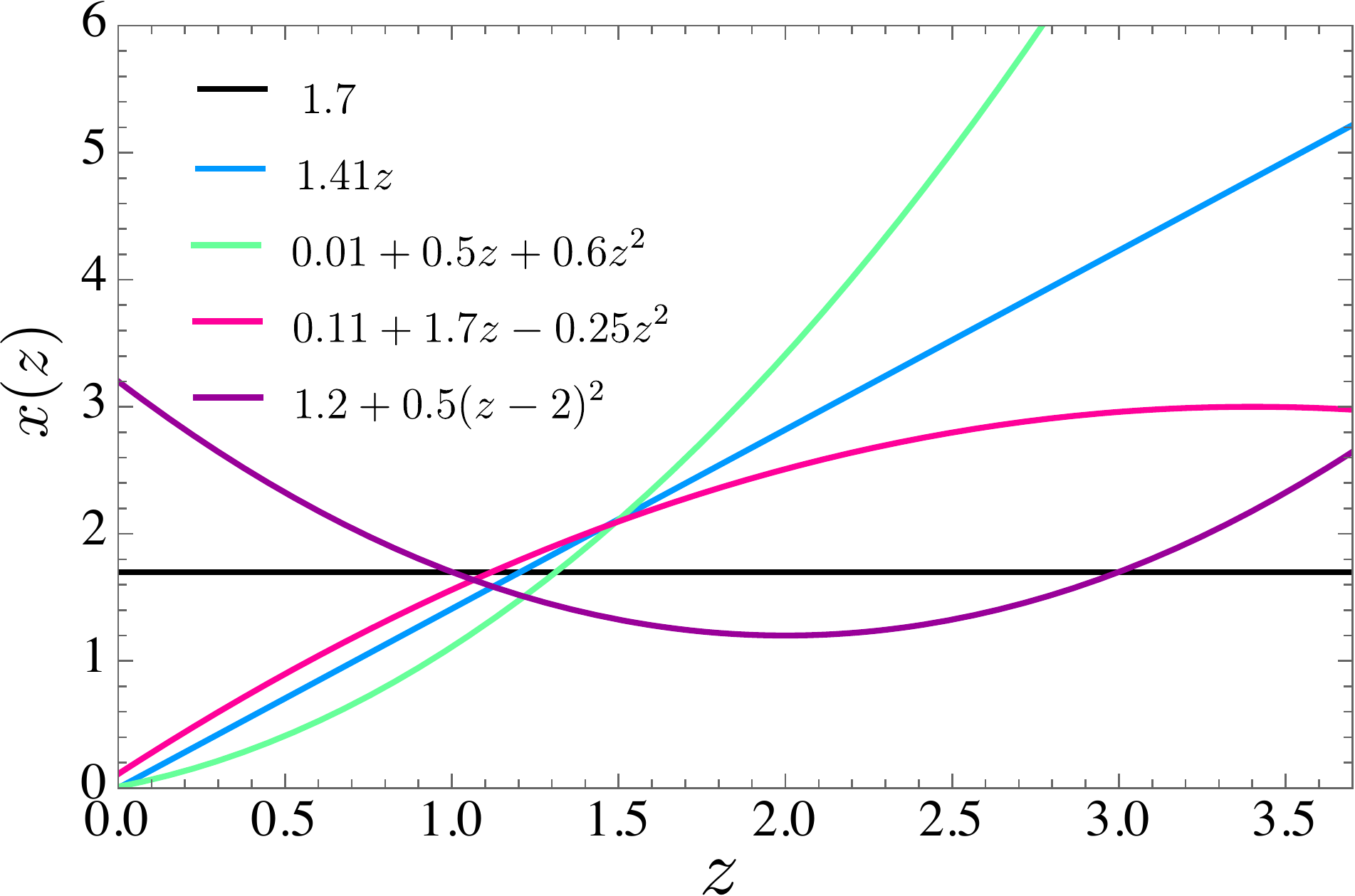}
    \caption{We present different functions for the magnification bias $x(z)$ under the constraint that $x\e{eff}=1.7$ according to \eqref{eq:xeff}. For any of these parameterization, $\mathcal{D}\e{mag}\in [-0.0090,-0.0046]$.}
    \label{fig:xz}
\end{figure}
The magnification part depends on the evolution of $x(z)$. In~\cite{Secrest:2020has}, this parameter has been measured from the whole catalog leading to
\begin{align}
x_{\rm eff}=  \int_0^\infty \dd z f(z)x(z)=1.7\, .  \label{eq:xeff}
\end{align}
The redshift evolution of $x(z)$ is not taken into account in the theoretical prediction from Eq.~\eqref{eq:dipEB}. Here we assume different redshift evolutions for $x(z)$, under the constraints that $x_{\rm eff}=1.7$, and we calculate how $\mathcal{D}_{\rm mag}$ depends on the evolution. The magnification bias measured in the eBOSS sample decreases with redshift below $z=2.2$ and then increases, as can be seen in Fig.~\ref{fig:magnification_bias}. We model this behaviour by a parabola (purple line in Fig.~\ref{fig:xz}). For this measurement, an absolute magnitude threshold $M_*=-25$ was chosen~\citep{Wang:2020ibf} (i.e.\ the same luminosity threshold $L_*$ for all redshifts). The CatWISE2020 sample has however a fixed flux threshold $S_*$, which corresponds to different luminosity thresholds at different redshifts. In particular, at high redshift, the luminosity threshold $L_*(z)$ associated with a fixed $S_*$ is larger than at low redshift. This means that there is a large number of sources that we cannot see at high redshift, and therefore the slope of the cumulative number of sources above $L_*$ is expected to be steeper at high redshift. In general we expect therefore $x(z)$ to increase with redshift for a sample which is flux limited. 
We consider various models, with $x(z)$ constant, increasing linearly or quadratically with redshift, as can be seen in Fig.~\ref{fig:xz}. With this, we find a range of possible values
\be
\mathcal{D}_{\rm mag}\in [-0.0090,-0.0046] \, , \label{eq:Expected_Dmag}
\ee
the more negative value corresponding to the parabola and the less negative value to the green curve. As before, the magnification dipole is negative, and decreases the overall amplitude of the total dipole. 

\begin{figure}
    \centering
    \includegraphics[width=0.47\textwidth]{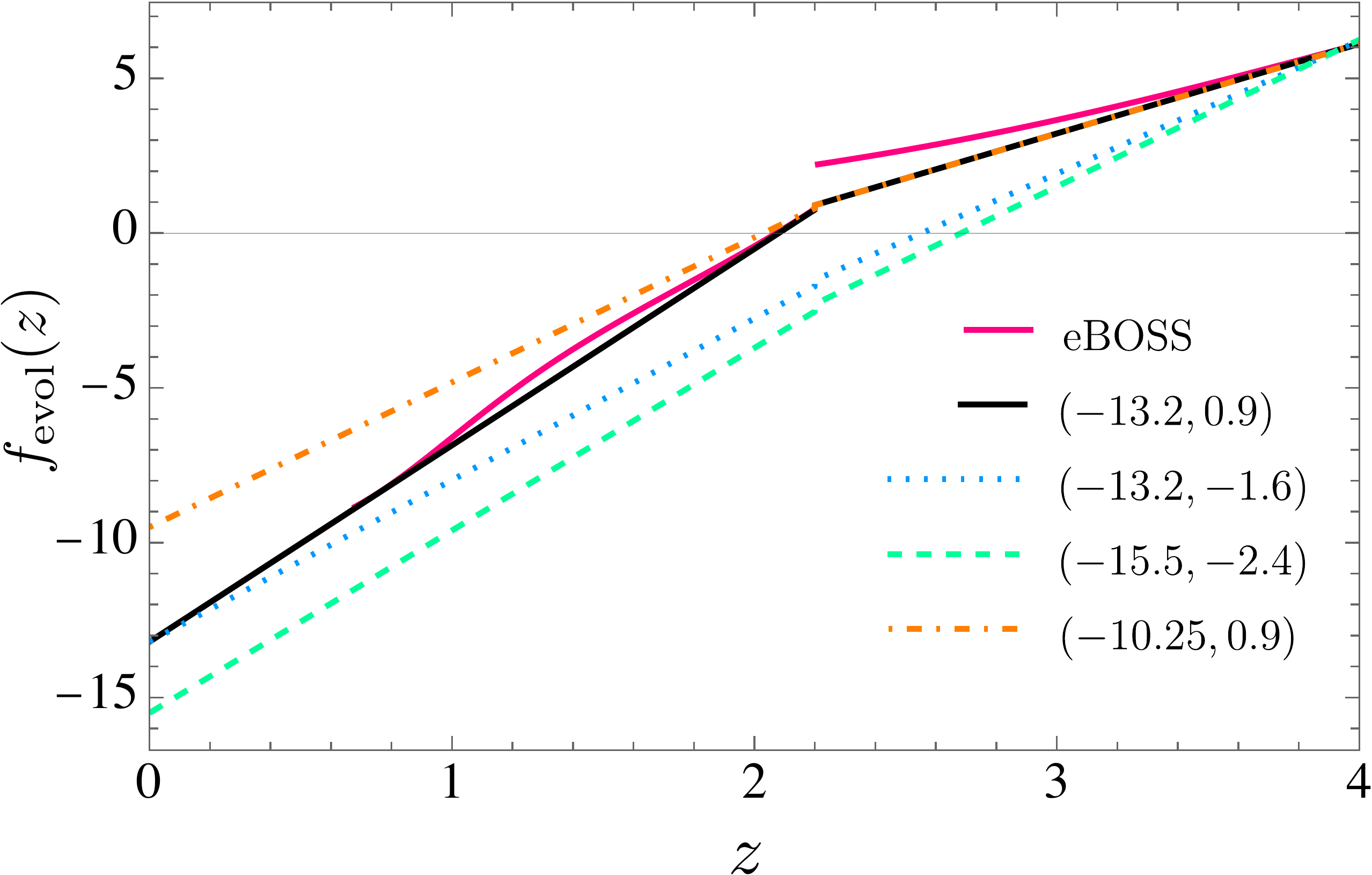}
    \caption{We show the evolution bias given by Eq.~\eqref{eq:fevolmodel}, with different values of the parameters $(f_0,f_1)$. The black line reproduces well the observed evolution bias measured from the eBOSS sample. The green long-dashed curve is an example for which the theoretical dipole agrees with the observed one at 2$\sigma$, whereas the blue short-dashed curve leads to an agreement at 3$\sigma$. The yellow dot-dashed curve is an example that would reproduce the theoretical prediction of Ellis and Baldwin, leading to a $4.9\sigma$ tension with observation.}
    \label{fig:fevolmodel}
\end{figure}

We can now infer the value that the evolution dipole should take in order to reproduce the dipole observed with the CatWISE2020 sample. For this we first compute the intrinsic clustering dipole and the shot noise dipole. As before, the clustering dipole can be calculated in $\Lambda$CDM and we find $\sqrt{\langle \mathcal{D}\e{int}^2\rangle }=0.0003$, assuming a bias of unity. This is 50 times smaller than $\mathcal{D}\e{obs}$ and in agreement with the findings of \cite{Secrest:2020has}. 

The CatWISE2020 sample contains $N=1'355'352$ sources and has a sky coverage $f_{\rm sky}=0.5$, \citep{Secrest:2020has}, leading to a shot noise contamination of $\sqrt{ \langle\mathcal{D}\e{sn}^2\rangle}\simeq 0.0018$. This leads to an expected standard deviation $\sigma_\mathcal{D}\simeq \sqrt{3\sigma_r^2} =\sqrt{\mathcal{D}\e{int}^2+\mathcal{D}\e{sn}^2} =0.0018$. This value can be compared with the result of~\cite{Secrest:2020has}. They perform simulations of their null hypothesis from which they obtain a distribution of dipole vectors. Out of $10^7$ simulations, only 5 have an amplitude larger than the observed value and a direction within the observed
angular distance from the CMB dipole. Approximating this distribution by a Gaussian, they quote a tension of 4.9$\sigma$. We use this to approximate the width of their distribution centered around the prediction $\mathcal{D}_{\rm EB}=0.0072$: $ \sigma_{\mathcal{D}}=|\mathcal{D}\e{obs}- \mathcal{D}\e{EB}|/4.9 = 0.0017$, which is very close to our theoretical value of 0.0018.
Since we want to infer the value that the evolution dipole should take to remove the observed tension, we use the "observational" error bars, $\sigma_{\mathcal{D}}=0.0017$, instead of the theoretical ones in the following.

Inserting $\sigma_{\mathcal{D}}$ in Eq.~\eqref{eq:tension}, we can find the value that $\langle \mathcal{D}\rangle$ should take for a given tension. We can then solve Eqs.\,\eqref{eq:D_Theory} and \eqref{eq:sigma_D_Theory} for $\mathcal{D}\e{kin}$ and $\sigma_r$. We find that the kinematic dipole should reach $\mathcal{D}_{\rm kin}= 0.0136$ (with $\sigma_r = 0.0017$) if we want the theoretical prediction to agree with the observed dipole at 1$\sigma$ ($\mathcal{D}_{\rm kin}=0.0118$, $\sigma_r=0.0017$ for 2$\sigma$ and $\mathcal{D}\e{kin}=0.0101$, $\sigma_r=0.0017$ for $3\sigma$).

Taking as an example for $x(z)$ the blue line from Fig.~\ref{fig:xz} (linear evolution) we have $\mathcal{D}_{\rm mag}=-0.0051$, meaning that an evolution dipole of  
\be
\mathcal{D}_{\rm evol}=\mathcal{D}_{\rm kin}-\mathcal{D}_{\rm cosmo}-\mathcal{D}_{\rm mag}=0.0121
\ee
would lead to an agreement at 1$\sigma$ ($\mathcal{D}\e{evol}=0.0103$ for 2$\sigma$ and $\mathcal{D}\e{evol}=0.0086$ for $3\sigma$). From Fig.~\ref{fig:fevol}, we see that the evolution bias evolves quasi-linearly with redshift, with a different slope before and after the pivot redshift $z_{\rm p}=2.2$, and that it vanishes around $z=2$. We take this behaviour as a starting point to define the following model for the evolution bias
\be
f_{\rm evol}(z)=\left\{\begin{array}{ll}
 f_0+0.45(f_1-f_0)z    &  {\rm if}\quad z\leq 2.2 \\
 2.2f_1-7.45+(3.4-0.56f_1)z   & {\rm if}\quad z>2.2\, ,
\end{array} \right.\label{eq:fevolmodel}
\ee
which changes slope at $z=2.2$, and reaches the eBOSS value at high redshift: $f_{\rm evol}(z=4)=6.1$. As can been seen on Fig.~\ref{fig:fevolmodel}, for $f_0=-13.2$ and $f_1=0.9$ the model (black line) reproduces relatively well eBOSS observations (red line). In this case, we find that $\mathcal{D}_{\rm evol}=0.0069$ leading to a $4.0\sigma$ tension with observations. If instead of the linear magnification bias, we use the quadratic one (green curve) in Fig.~\ref{fig:xz} the tension is reduced to 3.6$\sigma$. The tension is also reduced if the evolution bias is sightly more negative at low redshift than the eBOSS one. In Fig.~\ref{fig:fevolmodel}, we show two possible examples: the blue dashed line corresponds to an evolution dipole of $\mathcal{D}_{\rm evol}=0.0085$, i.e.\ for which the tension between the theoretical and observed dipole is of 3$\sigma$ and the green dashed line corresponds to a dipole of $\mathcal{D}_{\rm evol}=0.0103$ such that the theoretical prediction agrees at 2$\sigma$ with the observed dipole~\footnote{Note that for these results we use the linear magnification bias (blue curve in Fig.~\ref{fig:xz}). The tension would be even smaller if we would use instead the quadratic magnification bias (green curve).}.  We see that the behaviour of $f_{\rm evol}$ is not far from the one observed for the eBOSS sample. Since the eBOSS quasars were identified from the WISE survey~\citep{Ross:2020lqz,2015}, from which the CatWISE2020 catalog has been built, it is reasonable to think that the evolution bias is similar for the two catalogs. Nevertheless, it should be pointed out that small variations in the evolution bias parameter can lead to strong variations in $\mathcal{D}\e{evol}$.

We also calculate the value of the evolution dipole that we would need in order to reproduce the dipole predicted by Ellis and Baldwin, $\mathcal{D}_{\rm EB}=0.0072$. We find $\mathcal{D}_{\rm evol} = 0.0057$, corresponding to an $f_{\rm evol}$ which is larger (less negative) than the one extracted from eBOSS. We show such an example in Fig.~\ref{fig:fevolmodel} (dot-dashed yellow line). We see that the slope of this model is smaller than the one observed in eBOSS.

\section{Conclusion}
\label{sec:conclusion}

In this paper, we have derived theoretical expressions for the kinematic dipole of radio sources, explicitly accounting for the fact that the properties of the sources may evolve with redshift. We have presented two different calculations: in the first one the kinematic dipole is integrated over the distance of the sources, whereas in the second one it is integrated over their redshift. We have shown that the first derivation (integration over distance) agrees with the results of Ellis and Baldwin, if the spectral index of the sources and the magnification bias parameter are independent of redshift. However, if both of these quantities depend on redshift, Ellis and Baldwin's theoretical prediction is not fully representative of what is observed.

We have then shown that the second expression (integration over redshift) is mathematically equivalent to the first one, but that it has the advantage of being independent on the spectral index of the sources. Since this spectral index varies strongly from sources to sources (see for example Fig.\,2 of~\citet{Secrest:2020has}) and since we do not know if it exhibits a strong dependence on redshift, having a theoretical expression which does not rely on this quantity provides an important cross-check.

As a proof of concept on how this redshift theoretical expression can be used, we have applied it to the eBOSS quasar sample. This sample does not have enough sources to reliably measure the dipole. However, since the redshift of the quasars has been measured accurately, a measurement of the magnification bias and of the evolution bias have been performed in this sample~\citep{Wang:2020ibf}. We have used this to predict the amplitude of the kinematic dipole for the eBOSS quasars. We have then transposed these results to the CatWISE2020 sample, for which the dipole has been measured recently, showing a $4.9\sigma$ tension with respect to the expected dipole in $\Lambda$CDM~\citep{Secrest:2020has}. In particular we have determined which behaviour the evolution bias should have in order to reduce the tension to $2$
or $3\sigma$. Our results show that if the sources evolve at a rate which is one order of magnitude larger than the expansion rate (i.e.\ an $f_{\rm evol}$ of the order of -10), the tension with the CMB dipole disappears. Such a rate of evolution is similar to the measurement from the eBOSS sample.

Our work does not claim to solve the tension between the radio or quasar dipole and the CMB dipole, because we have no measurement of the evolution bias for the CatWISE2020 sample, which could be different from the one measured in eBOSS (for example, the different selection procedure in the two samples: absolute magnitude threshold versus flux threshold, could lead to a difference in the evolution bias). However, our results indicate two things: first that neglecting the redshift evolution of the sources properties (spectral index and magnification bias) may lead to an inaccurate theoretical prediction; and second that an alternative theoretical expression can be used, which relies on the evolution bias, and which is able to reduce the tension for reasonable evolution rates of the quasar population.

In the near future, the Dark Energy Spectroscopic Instrument (DESI) is expected to observe 2.4 million quasars over 14'000 square degrees, with spectroscopic redshifts~\citep{Farr:2020flh, DESI:2016fyo}. This will be an ideal catalog to measure the dipole, as well as the magnification bias and evolution bias of the quasars population, and to determine, through our new theoretical expression, if the tension with the CMB dipole is real or not.

\section*{Acknowledgements}

We would like to thank Subir Sarkar, Sebastian Von Hausegger, Tobias Nadolny, Giulia Cusin, Francesca Lepori, Chris Clarkson, Pierre Fleury, J\'er\'emie Francfort and Caroline Guandalin for useful discussions, Mike Wang for useful interactions and Mohamed Rameez for sharing their data. We also thank Ruth Durrer for useful comments on a first version of this work. C.D.\,was supported by a Swiss National Science Foundation (SNSF) Professorship grant (No.~170547). CB acknowledges funding from the Swiss National Science Foundation and from the European Research Council (ERC) under the European Union's Horizon 2020 research and innovation programme (Grant agreement No. 863929; project title
"Testing the law of gravity with novel large-scale structure observables").

\section*{Data availability}
The datasets were derived from sources in the public domain: \url{https://doi.org/10.5281/zenodo.4431089}.



\bibliographystyle{mnras}
\bibliography{references} 



\appendix

\section{Flux-luminosity relation}
\label{app:flux}

We derive here the relation between the flux density and luminosity density. Let us consider a source observed with flux density
\begin{align}
S(\nu_o)=\frac{\dd \mathcal{S}}{\dd \nu_o}=\frac{\dd n_{\nu_o} h \nu_o}{\dd \nu_o \dd \tau_o \dd A_o}\, ,   
\end{align}
where $h$ is the Planck constant. Here $\dd n_{\nu_o}$ denotes the number of photons received by the observer on a surface $\dd A_o$, during proper time $\dd \tau_o$, and at frequency $[\nu_o, \nu_o+\dd \nu_o]$.  The intrinsic luminosity density of the source is then given by
\begin{align}
 L(r,\bn,\nu_s)=\frac{\dd \mathcal{L}(r,\bn)}{\dd \nu_s}=\frac{4\pi \dd n_{\nu_s} h \nu_s}{\dd \nu_s \dd \tau_s \dd \Omega_s}\, ,  
\end{align}
where $\dd n_{\nu_s}$ denotes the number of photons emitted by the source in the fraction of solid angle $\dd \Omega_s/(4\pi)$ during proper time $\dd \tau_s$ at frequency $\nu_s=(1+z)\nu_o$. Using that $\dd n_{\nu_s}=\dd n_{\nu_o}$, i.e.\ all the photons emitted by the source in the solid angle $\dd \Omega_s/(4\pi)$ are received by the observer on the surface $\dd A_o$, we find
\begin{align}
 L(r,\bn,\nu_s)=4\pi \frac{d_L^2(r,\bn)}{1+z(r,\bn)}S(\nu_o)\, ,  \label{eq:luminosity}
\end{align}
where $d_L$ is the luminosity distance~\citep{Bonvin:2005ps}:
\be
d_L=(1+z)\sqrt{\frac{\dd A_o}{\dd \Omega_s}} = (1+z)^2 \sqrt{\frac{\dd A_s}{\dd \Omega_o}} = (1+z)^2 d_A\, .
\ee
Note that integrating Eq.~\eqref{eq:luminosity} over frequency we find the standard relation between flux and luminosity
\begin{align}
\mathcal{L}(r,\bn)&\equiv\int_0^\infty \dd\nu_s \frac{\dd \mathcal{L}(r,\bn)}{\dd \nu_s}\\
&= 4\pi\int_0^\infty \dd \nu_o (1+z)\frac{d_L^2}{(1+z)} \frac{\dd \mathcal{S}}{\dd \nu_o}=4\pi d_L^2 \mathcal{S}\nonumber\, .
\end{align}


\bsp	
\label{lastpage}
\end{document}